\documentclass[12pt,a4paper]{elsart}
\usepackage{epsfig}
\usepackage{graphics}
\begin{document}
\begin{frontmatter}
\title{Study of the Process $e^+e^- \to K^0_L K^0_S$
in the C.M.Energy Range \mbox{1.05~--~1.38 GeV} with CMD-2}
\author[BINP]{R.R.Akhmetshin},
\author[BINP,NSU]{V.M.Aulchenko},
\author[BINP]{V.Sh.Banzarov},
\author[BINP,NSU]{L.M.Barkov},
\author[BINP]{S.E.Baru},
\author[BINP]{N.S.Bashtovoy},
\author[BINP,NSU]{A.E.Bondar},
\author[BINP]{D.V.Bondarev},
\author[BINP]{A.V.Bragin},
\author[YALE]{S.K.Dhawan},
\author[BINP,NSU]{S.I.Eidelman},
\author[BINP,NSU]{D.A.Epifanov},
\author[BINP,NSU]{G.V.Fedotovich},
\author[BINP]{N.I.Gabyshev},
\author[BINP,NSU]{D.A.Gorbachev},
\author[BINP]{A.A.Grebeniuk},
\author[BINP,NSU]{D.N.Grigoriev},
\author[YALE]{V.W.Hughes},
\author[BINP,NSU]{F.V.Ignatov},
\author[BINP]{S.V.Karpov},
\author[BINP,NSU]{V.F.Kazanin},
\author[BINP,NSU]{B.I.Khazin},
\author[BINP,NSU]{I.A.Koop},
\author[BINP,NSU]{P.P.Krokovny},
%\author[BINP]{L.M.Kurdadze},
\author[BINP,NSU]{A.S.Kuzmin},
\author[BINP]{I.B.Logashenko},
\author[BINP]{P.A.Lukin\thanksref{someone}},
\author[BINP]{A.P.Lysenko},
\author[BINP]{K.Yu.Mikhailov},
\author[BINP,NSU]{I.N.Nesterenko},
\author[BINP]{V.S.Okhapkin},
\author[BINP,NSU]{A.V.Pak},
\author[BINP]{A.A.Polunin},
\author[BINP]{A.S.Popov},
\author[BU]{B.L.Roberts},
\author[BINP]{N.I.Root},
\author[BINP]{A.A.Ruban},
\author[BINP]{N.M.Ryskulov},
\author[BINP]{A.G.Shamov},
\author[BINP]{Yu.M.Shatunov},
\author[BINP,NSU]{B.A.Shwartz},
\author[BINP,NSU]{A.L.Sibidanov},
\author[BINP]{V.A.Sidorov},
\author[BINP]{A.N.Skrinsky},
%\author[BINP]{V.P.Smakhtin},
\author[BINP]{I.G.Snopkov},
\author[BINP,NSU]{E.P.Solodov},
\author[BINP]{P.Yu.Stepanov},
\author[BINP]{A.I.Sukhanov},
\author[PITT]{J.A.Thompson},
\author[BINP]{Yu.V.Yudin},
\author[BINP]{S.G.Zverev}

\address[BINP]{Budker Institute of Nuclear Physics, Novosibirsk,
630090, Russia}
\address[NSU]{Novosibirsk State University, Novosibirsk, 630090,
Russia}
\address[YALE]{Yale University, New Haven, CT 06511, USA}
\address[BU]{Boston University, Boston, MA 02215, USA}
\address[PITT]{University of Pittsburgh, Pittsburgh, PA, 15260, USA}
\thanks[someone]{contact person. e-mail:P.A.Lukin@inp.nsk.su}

\begin{abstract}
The process  $e^+e^- \to K^0_L K^0_S$
has been studied  with the CMD-2 detector       using about 950
events detected  in the center-of-mass
energy range from 1.05 to 1.38~GeV. The cross section
exceeds the expectation based on the contributions of the 
$\rho(770)$, $\omega(782)$ and $\phi(1020)$ mesons only. 
\end{abstract}
%%%%%
\end{frontmatter}
\section{Introduction}
\hspace*{\parindent} The investigation of the process $e^+e^-\to K^0_LK^0_S$
is important for a number of physics problems. 
%Studies of neutral kaon
%production provide valuable information about the K-meson structure. 
Since both I=0 and I=1 vector mesons can decay into a kaon pair,  
one can search for excitations of the $\rho(770), \omega(782)$ and 
$\phi(1020)$ by measuring  the cross section of the process in the energy
range above the $\phi(1020)$ meson~\cite{MANE81}. The isovector part
of the cross section of the process $e^+e^-\to K\bar{K}$ (both
$K^+K^-$ and $K^0_LK^0_S$ final states should be taken) 
can be related to the $\tau^- \to K^-K^0\nu_{\tau}$  decay by using
conservation of vector current \cite{cvc}. Assuming the hypothesis of 
factorization it can be also used to account for the production of 
kaon pairs in $B^- \to D^0 K^- K^0$ decays \cite{dru}.   
Finally, the process under study contributes to the total hadronic 
cross section of $e^+e^-$ annihilation, so that the values of its
cross section are used in the calculation of the hadronic contribution to 
the muon anomalous magnetic moment~\cite{g-21}. In view of the
increasing experimental accuracy in the measurement of this 
quantity \cite{g-22}, any significant contribution like that from the 
process $e^+e^-\to K^0_LK^0_S$ should be measured with 
adequate precision.

\par Earlier measurements of the cross section performed by the DM1
collaboration in Orsay~\cite{MANE81} and at OLYA and CMD detectors
in Novosibirsk~\cite{OLYA82,Solodov84} were based on small data
samples and had a  systematic accuracy of about 20\% or worse. 
Significant progress in the study of the process $e^+e^-\to K^0_LK^0_S$ was
achieved by the SND collaboration~\cite{SND99}
at the VEPP-2M collider~\cite{VEPP-2M}. The experiment was based on
integrated luminosity of 6.3~pb$^{-1}$. The systematic error of the
cross section was estimated to be 10\% around
1.04~GeV increasing to about 15\% at 1.38 GeV.
\par In this work we report on the measurement of the 
$e^+e^-\to K^0_LK^0_S$ cross section 
based on 5.7 pb$^{-1}$ of data collected with the CMD-2 Detector
\cite{CMD-2} at the VEPP-2M collider from 1.05 to 1.38~GeV.
The systematic uncertainty on the
cross section was  about 5\% below 1.09~GeV and
increased to 10\% at 1.38~GeV.
\section{Detector and experiment}
\hspace*{\parindent} The CMD-2 detector % shown in Fig.~\ref{p:cmd2}
 has been described in detail
elsewhere~\cite{CMD-2}. 
%The longitudal and transversal
%detector cross sections are shown in Fig.~\ref{p:cmd2}.
Its tracking system  consists of the cylindrical
drift chamber (DC)~\cite{DC} surrounding the  interaction point,
and proportional Z-chamber (ZC)~\cite{ZC} for precise polar angle
measurement, both also used for the trigger. Both chambers are inside a
thin (0.38~$X_0$) superconducting solenoid~\cite{HFIELD} with a field of 1~T.
The barrel electromagnetic calorimeter~\cite{CsI} is
placed outside the solenoid and consists of
892 CsI crystals. The muon-range system~\cite{MU} of
the detector, also located outside the solenoid, is based on streamer tubes.
The endcap electromagnetic calorimeter~\cite{BGO} based
on BGO crystals makes the detector almost hermetic for photons.
%\begin{figure}
%{\centering \begin{tabular}{cc}
%\resizebox*{0.4\textwidth}{!}{\includegraphics{cmd2_l.eps}} &
%\resizebox*{0.4\textwidth}{!}{\includegraphics{cmd2_t.eps}} \\
%\end{tabular}\par}
%\caption{\label{p:cmd2}Longitudal and transverse cross sections of CMD-2
%detector: 1.~---~vacuum pipe; 2~---~drift chamber; 3~---~Z-chamber;
%4~---~superconducting solenoid; 5~---~compensating solenoid;
%6~---~electromagnetic BGO endcap calorimeter; 7~---~electromagnetic CsI barrel
%calorimeter; 8~---~range system; 9~---~magnet yoke; 10~---~quadrupole lenses.}
%\end{figure}
\par The data sample used in the analysis was collected in two scans of
the center-of-mass energy range 1.05~--~1.38~GeV. In the first scan
the  beam energy was increased
from 530~MeV to 690~MeV with a 10~MeV step, while in the second one
it was decreased from 685~MeV to 525~MeV with the same energy
step.
\section{Data analysis}
\hspace*{\parindent} Events of the process $e^+e^-\to K^0_L K^0_S$
were detected by using a subsequent decay of the $K^0_S$ meson into a 
pair of charged pions.
\par
\begin{figure}
\includegraphics[width=0.9\textwidth]{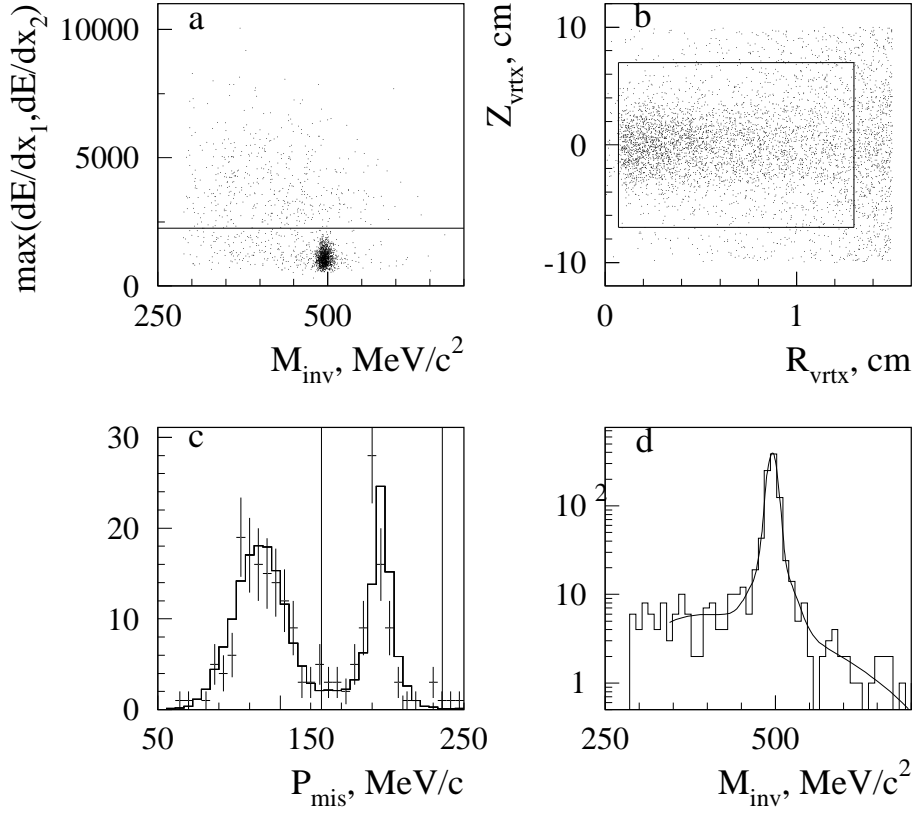}
\caption{\label{cuts} Distributions used for the selection
of  $e^+e^- \to K^0_L K^0_S$ events: a. Particle ionization losses 
versus the invariant mass of two tracks; b. Z-coordinate of the 
vertex versus the distance
from the beam point to the vertex in the $R-\varphi$ plane; 
c. Missing momentum (the histogram is simulation, the points are
experimental events); d. Invariant mass of two tracks.}
\end{figure}

 The following selection criteria were used:
\begin{itemize}
\item There are two oppositely charged tracks from the vertex closest to the
beam. Track momenta, assuming that tracks are pions, satisfy the 
conditions:
$$
P^{min}_{\pi} - 20 < P_{1,2} < P^{max}_{\pi} + 20,
$$
where $P^{min}_{\pi},P^{max}_{\pi}$ are minimum and maximum kinematically
possible momenta of a pion from the $K^0_S\to\pi^+\pi^-$ decay in MeV/c.
\item The track polar angles are:
$$
0.95 < \theta_{1,2} < \pi - 0.95.
$$
\item The maximum of the track ionization losses is
$$
%max(\frac{dE}{dx_1},\frac{dE}{dx_2}) < 2.2\frac{dE}{dx_{MIP}},
max((\frac{dE}{dx})_1,(\frac{dE}{dx})_2) < 2.2 (\frac{dE}{dx})_{MIP},
$$
where $(\frac{dE}{dx})_{MIP}$ is the ionization loss of a minimum ionizing
particle (see Fig.~\ref{cuts}a for events from the energy range 
$1.05 < \sqrt {s} < 1.09$ GeV). 
The cut is used to remove $K^+K^-$ events
as well as events of ``beam-wall'' interactions.
\item The acollinearity angle for pions in the R-$\varphi$ plane is:
$$
0.2 < |\pi - |\varphi_1 - \varphi_2|| < 3.0,
$$
and the space angle between pion tracks is
$$
\psi > 0.5.
$$
By these criteria events with collinear particles and particles 
going close to each other were rejected.
\item The Z-coordinate of the vertex is less than 7.0 cm and the 
radius of the vertex in the R-$\varphi$ plane ($R_{vrtx}$) satisfies 
the condition:
$$
0.07 < R_{vrtx} < 1.3\mbox{ cm,}
$$
as shown in Fig.~\ref{cuts}b for the energy range 
$~1.05 < \sqrt {s} < 1.09$ GeV.
\end{itemize}
\par In the energy region above the $\phi(1020)$, production of a neutral 
kaon pair is often accompanied by emission of a hard photon by initial
electrons (``return to resonance'' effect). The distribution of the
missing momentum of two tracks defined as 
$P_{mis}=|{\vec{P}}_1+{\vec{P}}_2|$ 
is shown in Fig.~\ref{cuts}c for $\sqrt {s} = 1.07$ GeV. In
agreement with the Monte Carlo simulation, the left peak in
Fig.~\ref{cuts}c 
corresponds to ``return to resonance'' events, 
while the right one describes events without hard photon emission. 
In the present work events with ``return to resonance'' were excluded 
from analysis by the requirement:
$$
  \left| P_{mis} - \sqrt{E^2_{beam}-m^{2}_{K^0}} \right| < 40~\mbox{MeV/c,}
$$
where 40 MeV/c corresponds to five standard deviations of the 
experimental resolution for the $P_{mis}$ value.
The selection criteria are shown in Fig.~\ref{cuts}c by the 
vertical lines.
\par The number of events was determined from a fit of the
invariant mass distribution with a sum of two Gaussians describing 
the signal and a smooth function describing background. 
%To determine parameters of the background function 
%as well as variances of the signal Gaussians, the following procedure 
%was used. 
For the fitting procedure, the data sample was subdivided into three energy
bins: 1.05--1.19, 1.20--1.29 and 1.30--1.38 GeV. The approximation 
above was performed in each of the three bins and the parameters obtained
%parameters of the fitting functions were later fixed during the 
were later fixed during the approximation at each energy point within 
the corresponding bin.
An example of such an approximation in the energy range
1.05--1.19~GeV is shown in Fig.~\ref{cuts}d.
\par After background subtraction and application of  the cuts
described above, 948$\pm$33 $K^0_LK^0_S$ events were selected.
\par At each energy point the cross section is determined from
the following formula:
$$
\sigma = \frac{N}
{L\varepsilon(1+\delta_{rad})},
%\varepsilon_{rec}\varepsilon_{trig}\varepsilon_{geom}L(1+\delta_{rad})},
$$
where $N$ is the number of selected events, $\varepsilon$ is the 
detection efficiency, $L$ is the integrated luminosity 
determined from events of Bhabha scattering at large
angles~\cite{lum}, and $(1+\delta_{rad})$ is a radiative 
correction due to initial state radiation~\cite{rad}.
% and including the effects of both leptonic and 
%hadronic vacuum polarization.
The detection efficiency 
$\varepsilon~=~\varepsilon_{rec}\varepsilon_{trig}\varepsilon_{geom}$,
where $\varepsilon_{rec}$ is the reconstruction
efficiency, $\varepsilon_{trig}$ is the trigger efficiency, and
$\varepsilon_{geom}$ is the acceptance.
\par The reconstruction efficiency ($\varepsilon_{rec}$) was
determined from the experimental data~\cite{formf}.
% using the procedure described in~\cite{formf}. 
To this end ``test'' events, in which a $K^0_L$ produced a cluster 
in the CsI calorimeter, were selected. Using the angles of this cluster
as well as the angles of the clusters produced by pions from the
$K^0_S \to \pi^+\pi^-$ decay and requiring one track in DC, one 
obtains a clean sample of events for efficiency
determination. Analysis shows that the reconstruction efficiency is 
energy independent. 
%The same procedure was applied to simulation and 
%also gave an energy independent reconstruction efficiency. The 
%efficiency values in the data sample and simulation are slightly
%different, therefore a
%correction for this difference was applied to the experimental
%reconstruction efficiency. 
A similar procedure was used to determine
the trigger efficiency ($\varepsilon_{trig}$)~\cite{formf}. The acceptance
($\varepsilon_{geom}$) was determined from Monte Carlo simulation. 
%The following values of 
%the efficiency averaged over energy were obtained:
%\begin{eqnarray}
%\varepsilon_{rec} & = & 0.955\pm 0.018\nonumber,\\
%\varepsilon_{trig}& = & 0.981\pm 0.004\nonumber,\\
%\varepsilon_{geom}& = & 0.130\pm 0.001\nonumber.
%\end{eqnarray}
\par The beam energy at each point was evaluated from the value of the
magnetic field in the dipole magnets~\cite{energy}.
The systematic uncertainty was estimated to be
$\frac{\Delta E}{E} = 4\cdot 10^{-4}$ from the
analysis of the long-term stability of energy~\cite{lum}. The
number of events, integrated luminosity, detection efficiency, radiative
correction  and cross section at each energy point are
listed in Table~\ref{tab:cross}. It is this cross section (the
``dressed'' one) that should be used in the approximation of the
energy dependence with resonances. For applications to various dispersion 
integrals such as the leading order hadronic contribution 
to the muon anomalous magnetic moment, one should use the ``bare''
cross section. Following the procedure in Ref.~\cite{pipi}, the latter 
is obtained from the ``dressed'' one by 
multiplying it by the vacuum polarization factor $|1~-~\Pi(s)|^2$,
where $\Pi(s)$ is the photon polarization operator calculated  
taking into account the effects of both leptonic and hadronic vacuum 
polarization.
\begin{table}
\caption{The c.m. energy, number of events, 
integrated luminosity, detection efficiency,
radiative correction, ``dressed'' cross section, vacuum polarization
factor and ``bare'' cross section of the process $e^+e^-\to K^0_L
K^0_S$. The first error in the cross section value is  statistical and
the second is systematic one.}
\vspace*{0.5cm}
\begin{center}
\hspace*{-20mm}\begin{tabular}{cccccccc} 
\hline
$\sqrt{s}$, GeV & N & $\mathcal{L}$, nb$^{-1}$& $\varepsilon$ &
$(1+\delta_{rad})$ &$\sigma$, nb & $|1-\Pi(s)|^2$ & $\sigma^0$, nb\\ 
\hline
   1.050 &310.9$\pm$17.8
&117.7$\pm$0.7&0.151&0.996&17.56$\pm$1.00$\pm$
1.97&0.948&16.65$\pm$0.95$\pm$1.87 \\
   1.060 &124.1$\pm$11.3
&76.8$\pm$0.6&0.146&0.950&11.65$\pm$1.06$\pm$
0.64&0.953&11.10$\pm$0.97$\pm$0.61 \\
   1.070 & 76.1$\pm$8.9  &81.8$\pm$0.5&0.141&0.926&7.13$\pm$0.84$\pm$
0.35&0.956& 6.82$\pm$0.80$\pm$0.33 \\
   1.080 & 39.3$\pm$6.6  &59.6$\pm$0.5&0.136&0.914&5.30$\pm$0.90$\pm$
0.26&0.958& 5.08$\pm$0.86$\pm$0.25 \\
   1.090 & 53.0$\pm$7.5  &85.2$\pm$1.3&0.132&0.909&5.18$\pm$0.73$\pm$
0.25&0.959
   & 4.97$\pm$0.70$\pm$0.24 \\
   1.100 & 29.0$\pm$5.5  &57.5$\pm$0.5&0.128&0.906&4.35$\pm$0.81$\pm$
0.21&0.961
   & 4.19$\pm$0.78$\pm$0.20 \\
   1.110 & 33.5$\pm$6.1  &83.3$\pm$0.5&0.124&0.905&3.58$\pm$0.65$\pm$
0.29&0.962
   & 3.44$\pm$0.62$\pm$0.28 \\
   1.120 & 18.2$^{+4.9}_{-4.2}$&59.4$\pm$0.7&0.120&0.905&2.82$^{+0.75}_{-0.65}\pm$0.23&0.963
   & 2.71$^{+0.72}_{-0.62}\pm$0.22 \\
   1.130 & 14.7$^{+4.2}_{-3.6}$&62.8$\pm$0.5&0.116&0.905&2.23$^{+0.63}_{-0.55}\pm$0.18&0.963
   & 2.15$^{+0.61}_{-0.53}\pm$0.18 \\
   1.140 & 25.5$^{+5.7}_{-5.0}$&87.4$\pm$0.5&0.113&0.905&2.85$^{+0.61}_{-0.54}\pm$0.23&0.964
   & 2.75$^{+0.59}_{-0.52}\pm$0.22 \\
   1.150 & 17.8$^{+4.8}_{-4.1}$&63.2$\pm$0.5&0.109&0.906&2.85$^{+0.74}_{-0.62}\pm$0.23&0.965
   & 2.75$^{+0.71}_{-0.60}\pm$0.21 \\
   1.160 & 17.2$^{+4.8}_{-4.1}$&113.0$\pm$0.7&0.106&0.906&1.58$^{+0.44}_{-0.37}\pm$0.14&0.965
   & 1.52$^{+0.42}_{-0.36}\pm$0.13 \\
   1.170 & 13.9$^{+4.2}_{-3.5}$&71.8$\pm$0.5&0.103&0.907&2.07$^{+0.59}_{-0.50}\pm$0.17&0.966
   & 2.00$^{+0.57}_{-0.48}\pm$0.16 \\
   1.180 & 15.1$^{+4.6}_{-3.9}$&114.9$\pm$0.7&0.100&0.907&1.45$^{+0.42}_{-0.36}\pm$0.12&0.966
   & 1.40$^{+0.41}_{-0.35}\pm$0.12 \\
   1.190 & 20.9$^{+5.3}_{-4.6}$&132.5$\pm$1.1&0.097&0.908&1.79$^{+0.44}_{-0.37}\pm$0.15&0.966
   & 1.72$^{+0.42}_{-0.36}\pm$0.14 \\
   1.205 & 19.7$^{+5.6}_{-4.9}$ &270.1$\pm$1.1&0.094&0.909&0.85$^{+0.23}_{-0.21}\pm$0.07&0.967
   & 0.82$^{+0.23}_{-0.21}\pm$0.07 \\
   1.225 & 26.9$^{+6.0}_{-5.3}$&286.7$\pm$1.5&0.089&0.910&1.16$^{+0.24}_{-0.21}\pm$0.11&0.967
   & 1.12$^{+0.24}_{-0.21}\pm$0.11 \\
   1.251 & 21.3$^{+5.7}_{-5.3}$&492.3$\pm$2.3&0.083&0.912&0.57$^{+0.14}_{-0.14}\pm$0.05&0.968
   & 0.55$^{+0.14}_{-0.14}\pm$0.05 \\
   1.275 & 18.7$^{+5.8}_{-5.0}$&468.6$\pm$1.5&0.079&0.913&0.55$^{+0.17}_{-0.14}\pm$0.05&0.968
   & 0.53$^{+0.17}_{-0.14}\pm$0.05 \\
   1.296 & 20.2$^{+6.0}_{-5.2}$&589.2$\pm$2.7&0.075&0.915&0.50$^{+0.14}_{-0.12}\pm$0.05&0.968
   & 0.48$^{+0.14}_{-0.12}\pm$0.05 \\
   1.325 &11.2$^{+6.2}_{-5.4}$&994.9$\pm$3.1&0.071&0.916&0.17$
^{+0.09}_{-0.08}\pm$0.02&0.968
   & 0.16$^{+0.09}_{-0.08}\pm$0.02 \\
   1.368 & 20.7$^{+6.6}_{-5.9}$ &1338.7$\pm$3.4&0.066&0.919&0.25$
^{+0.08}_{-0.07}\pm$0.02&0.968
   & 0.24$^{+0.08}_{-0.07}\pm$0.02 \\ 
\hline
\end{tabular}
\end{center}
\label{tab:cross}
\end{table}

\par
\begin{figure}
\includegraphics[width=0.9\textwidth]{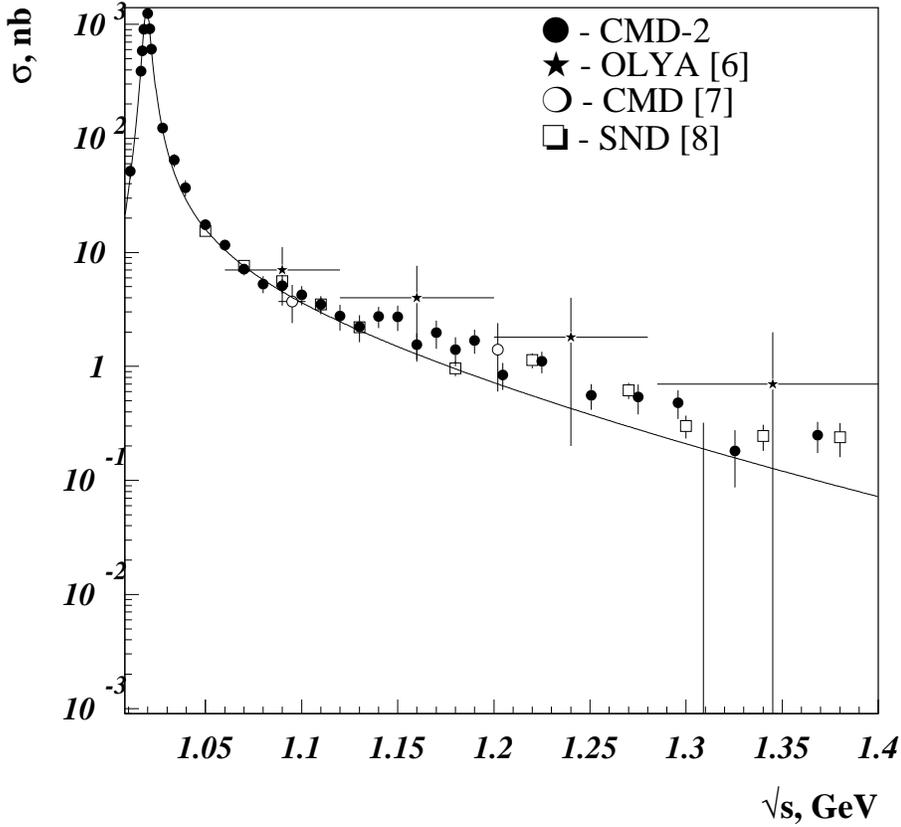}
\caption{\label{fit} The cross section of the process $e^+e^-\to
K^0_LK^0_S$ in the energy range $\sqrt {s}$~=~1.05~--~1.38~GeV,
measured in different experiments. The curve is the
Vector Dominance Model prediction with the contributions of the
$\rho(770)$, $\omega(782)$ and $\phi(1020)$ mesons.}
\end{figure}

 Figure~\ref{fit} shows the energy dependence of the cross section
obtained in this work together with the results of our study of the
$\phi\to K^0_L K^0_S$ decay from~\cite{exp:phi} (the data in
the 1.00~--~1.04~GeV energy range). Also shown are the results of the
previous experiments that studied the process $e^+e^-\to K^0_LK^0_S$ 
above the $\phi$ meson region. Good agreement between the results of all
measurements is observed.
\par
\begin{figure}
\includegraphics[width=0.9\textwidth]{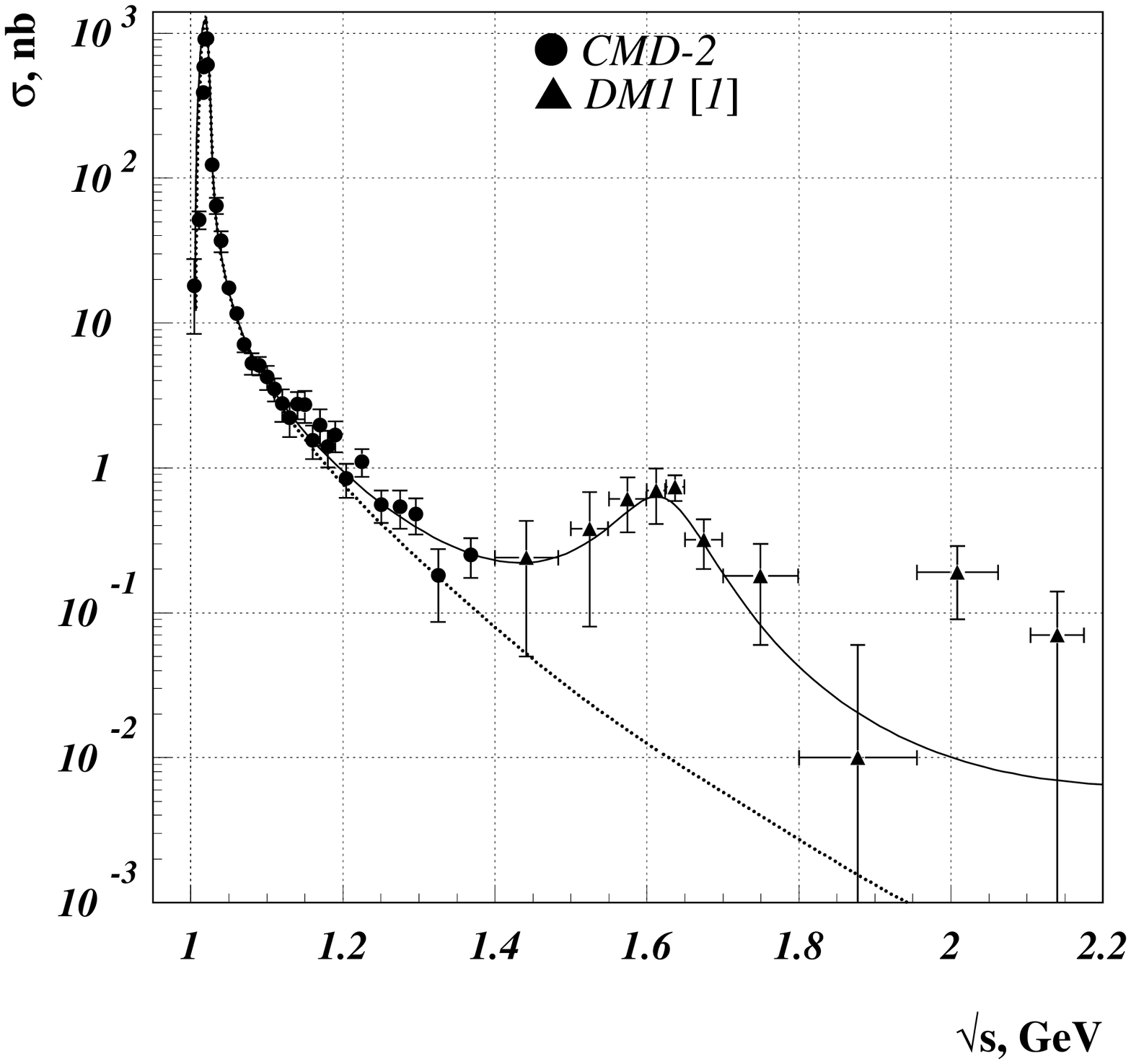}
\caption{\label{inter} The cross section of the process $e^+e^-\to
K^0_LK^0_S$ in the energy range $\sqrt {s}$~=~1.05~--~2.2~GeV compared
to the Vector Dominance Model predictions: $\rho(770)$, $\omega(782)$, 
$\phi(1020)$ mesons (the dotted curve) and $\rho(770)$, $\omega(782)$, 
$\phi(1020)$ and X (the solid curve).}
\end{figure}

 The energy dependence of the cross section was approximated in 
the frame of Vector Dominance Model~\cite{VDM} with the contributions 
of the $\rho(770)$, $\omega(782)$ and
$\phi(1020)$ mesons according to the following SU(3) based formula:
\begin{eqnarray}
\sigma(s) & = & \frac{1}{s^{5/2}}\cdot\frac{q^3(s)}{q^3(m^2_{\phi})}\cdot
\left|-\frac{\Gamma_{\phi}m^3_{\phi}\sqrt{\sigma(\phi\to K^0_LK^0_S)m_{\phi}}}
{D_{\phi}(s)} -\right.\nonumber\\
 &-&\frac{\sqrt{\Gamma_{\phi}\Gamma_{\omega}m^2_{\phi}m^3_{\omega}6\pi
 B(\omega\to e^+e^-)B(\phi\to K^0_LK^0_S)}}{D_{\omega}(s)} + \nonumber\\  
 &+&\left. \frac{\sqrt{\Gamma_{\phi}\Gamma_{\rho}m^2_{\phi}m^3_{\rho}6\pi
  B(\rho\to e^+e^-)B(\phi\to K^0_LK^0_S)}}{D_{\rho}(s)}\right|^2\nonumber,
\end{eqnarray}
where 
$\sigma(\phi\to K^0_LK^0_S) = 12\pi B(\phi\to e^+e^-)B(\phi\to
K^0_LK^0_S)/m^2_{\phi}$ is the only free parameter, determined from the fit,
$q(s) = \sqrt{s/4 - m^2_{K^0}}$ is the neutral kaon momentum, and
$D_{V}(s) = m^2_V-s+\imath\sqrt{s}\Gamma_V(s)$.
The energy dependence of the total width 
for a meson $V$ was chosen as in~\cite{width}. Masses, total 
widths and branching ratios of the resonances were taken from~\cite{pdg}. 
The following 
value of the cross section at the $\phi$ meson peak
was obtained from the fit:
\begin{eqnarray}
\sigma(\phi\to K^0_LK^0_S) & = & (1376\pm 6\pm 23)\mbox{ nb},\nonumber\\
\chi^2/n.d.f               & = & 94.64/56 = 1.69.\nonumber
\end{eqnarray}
The value of the peak cross section is exactly equal
to that from our $\phi\to K^0_LK^0_S$ study~\cite{exp:phi}. The 
relatively large value of the $\chi^2$ arises from the energy range 
above 1.13 GeV where, 
%As can be seen from Fig.~\ref{fit}, there is a discrepance between 
%the experimental data and the approximation curve in the energy region 
%$E_{c.m.} > $ 1130~MeV. 
as seen from Fig.~\ref{fit}, most of the
experimental points lie above the approximation curve. 
One of the possible explanations for the observed excess could be
higher resonances contributing to the production of pairs of neutral kaons. 
%This assumption is confirmed by studying the process 
%$e^+e^-\to K^0_LK^0_S$ in the energy range 
%1.4~--~2.18~GeV with the DM1 detector~\cite{MANE81}. 
This assumption is confirmed by the combined analysis of our results
with those from the DM1 detector obtained in the energy range 
1.4~--~2.18~GeV~\cite{MANE81}, see Fig.~\ref{inter}. 
%The result 
%of the DM1 experiment is shown in Fig.~\ref{inter}
%by triangles. The result, obtained in the present work, is shown in 
%Fig.~\ref{inter} by points. 
To describe the energy dependence of the cross section 
at $\sqrt {s} \sim$ 1.6~GeV, the following amplitude was added to the
amplitudes of the $\rho(770)$, $\omega(782)$ and $\phi(1020)$ mesons:
$$
A_X = \frac{\sqrt{\frac{m^7_X\Gamma^2_X\sigma(X\to K^0_LK^0_S)q^3(m^2_{\phi})}
{q^3(m^2_X)}}}{m^2_X-s+\imath\sqrt{s}\Gamma_X}\cdot e^{\imath\delta_X}.
$$
The parameters $\sigma(X\to K^0_LK^0_S)$, $M_X$, $\Gamma_X$,$\delta_X$ 
as well as the cross section at the $\phi$ meson peak
were obtained from the approximation:
\begin{eqnarray}
\sigma(\phi\to K^0_LK^0_S)& = & (1375\pm 6\pm 23)\mbox{ nb},\nonumber\\
\sigma(X\to K^0_LK^0_S)   & = & 0.73\pm 0.33\mbox{ nb},\nonumber\\
M_X                       & = & 1623\pm 20\mbox{ MeV/c}^2,\nonumber\\
\Gamma_X                  & = & 139\pm 60\mbox{ MeV},\nonumber\\
\delta_X                  & = & 160^{\circ}\pm 42^{\circ},\nonumber\\
\chi^2/n.d.f             & = & 56.42/62 = 0.91.\nonumber
\end{eqnarray}
The value of the cross section at the $\phi$ meson peak changes from 
1376 nb to 1375 nb after the contribution of a higher resonance is 
taken into account.
This change agrees with the estimation of the systematic
uncertainty of the cross section (of about 1.5 nb) due to the 
model dependence of the cross section value~\cite{disser}.
The approximation is shown in Fig.~\ref{inter} by the solid line.
The theoretical curve describing the contributions of the $\rho(770)$,
$\omega(782)$ and $\phi(1020)$ mesons only, is shown in Fig.~\ref{inter} by
the dotted line. One can see that after adding the amplitude
$A_X$,  the quality of the approximation of the experimental data is
much better. The values obtained for the mass and width of the X state are
consistent with those of the $\phi(1680)$ meson \cite{pdg}. However, 
as noted above, both isovector and isoscalar states could contribute
to the cross section of the neutral kaon production in the energy
range 1.0~--~2.0~GeV. To identify unambiguously the nature of the
observed enhancement as well as to determine the relative weights of the
I=0 and I=1 final states,
one should simultaneously study both $K^0_LK^0_S$ and
$K^+K^-$ final states with a significantly higher data sample, 
particularly in the energy range above 1.4~GeV.
The detailed investigation of other final states of $e^+e^-$
annihilation in this energy range will be also needed to shed light on
the spectroscopy of the light quark resonances
in the vector sector.
Such studies will be possible at the VEPP-2000 collider~\cite{vep2000} 
now under construction at the Budker Institute of Nuclear Physics.

\begin{table}
\caption{Main sources of the systematic errors}
%\label{tab:syst}
\begin{tabular}{lc}
\hline
Source & Contribution,\% \\
\hline
Selection criteria & 3.5--6 \\
Background subtraction & 2--8 \\
Luminosity  & 2 \\
Detection efficiency  & 2 \\
Radiative corrections & 1 \\
\hline
Total       & 5~--~10 \\
\hline
\end{tabular}
\label{tab:syst}
\end{table}
The main sources of  the systematic uncertainties are listed
in Table~\ref{tab:syst}. 
The uncertainty caused by the selection criteria was estimated by
varying the cuts for pion momenta, acollinearity
angle $|\Delta{\varphi}|$ and Z-coordinate of the vertex by one standard 
deviation. It showed that the cross section changed by 3.5\% below 1.1 GeV
and by 6\% above this energy. To estimate the uncertainty due to the 
background shape, the invariant mass distribution of two tracks was 
approximated in a narrow mass range from 400 to 600 MeV/c$^2$, the 
parameters of the function describing background were determined and 
the obtained number of signal events was compared to that after the 
standard approximation in the invariant mass range from 340 to 700
MeV/c$^2$. The variation of the cross section was 2\% in the energy
range 1.05--1.09 GeV smoothly rising to 8\% in the higher energy range.  
The systematic error of the luminosity is caused by the radiative
corrections to the Bhabha scattering cross section as well as
selection of Bhabha events~\cite{lum}. The uncertainty of the
detection efficiency is dominated by the systematic error in the 
reconstruction efficiency and was estimated from the statistical 
error of the approximating constant in experiment~\cite{formf}. The 
uncertainty on the radiative correction comes from the
accuracy of the theoretical formulae used in the calculation (1\%) 
as well as from the missing momentum resolution. 
%The former effect changes the value of the radiative correction by 
%0.3\% above 1300 MeV and by less than 0.1\% at other energies. 
To take into account the influence of the latter effect, selection criteria
using this parameter were varied by one standard deviation. The
resulting change was significant at 1.05 and 1.06 GeV only where it
was 10\% and 2.5\% respectively whereas at all other energies
it was negligible.
The overall systematic uncertainty  of the cross section is obtained
by adding the individual contributions in quadrature. It grows with 
energy from 5\% in the energy range $\sqrt {s}$ = 1.05--1.09~GeV
to 10\% in the energy range  $\sqrt {s}$ = 1.27--1.38~GeV.

\section{Conclusion}
\hspace*{\parindent} Using 948$\pm$33 reconstructed events
detected by the CMD-2, the cross section of
the process $e^+e^-\to K^0_LK^0_S$ was determined
in the energy range from 1.05 to 1.38~GeV. This is the most precise
measurement of this cross section by now. It is shown that 
in the energy range $\sqrt {s} > 1.13$~GeV 
the obtained energy dependence of the cross section could not be
explained by the Vector Dominance Model with the contributions of the 
$\rho(770)$, $\omega(782)$ and $\phi(1020)$ mesons only, so that 
higher resonances should be taken into account. The addition of an
amplitude corresponding to a resonance with mass and width close to
those of the $\phi(1680)$ meson and interfering with the 
$\rho(770)$, $\omega(782)$ and $\phi(1020)$ mesons substantially
improves the description of the observed energy dependence.
To obtain the detailed information about the spectroscopy of 
higher resonances, new experiments are needed which will be 
performed at the VEPP-2000 collider.
\section{Acknowledgements}
\hspace*{\parindent} The authors are grateful to the staff of VEPP-2M for
excellent collider performance, to all engineers and technicians who
participated in the design, commissioning and operation of CMD-2.
We acknowledge useful constructive discussions with A.A.~Kozhevnikov
and A.I.~Milstein.
\par This work is supported by grants INTAS 99-00037, DOE DEF0291ER40646,
Integration A0100, NSF PHY 9722600, NSF PHY 0100468, RFBR 98-02-17851,
RFBR 02-02-16126a.

\end{document}